\documentclass[aps,prd,reprint,twocolumn,superscriptaddress,showpacs]{revtex4-1}
\usepackage{amsfonts}
\usepackage{mathrsfs}
\usepackage{amsmath}
\usepackage{color}
\usepackage{natbib}
\usepackage{graphicx}
\usepackage{bm}
\usepackage{amssymb}
\usepackage{xspace}
\usepackage{epstopdf}
\usepackage{dcolumn}
\usepackage{longtable}
\usepackage{multirow}
\usepackage[colorlinks=true, letterpaper=true, pdfstartview=FitV, linkcolor=blue, citecolor=blue, urlcolor=blue]{hyperref}
\usepackage{float}

\makeatletter

\newcommand{\Rmnum}[1]{\expandafter\@slowromancap\romannumeral #1@}
\makeatother

\begin{document}
\title{Topological nodal lines and hybrid Weyl nodes in YCoC$_2$}

\author{Yuanfeng Xu}
\email{Y. Gu and Y. Xu contributed equally to this work.}
\affiliation{Beijing National Laboratory for Condensed Matter Physics, and Institute of Physics, Chinese Academy of Sciences, Beijing 100190, China}
\affiliation{University of Chinese Academy of Sciences, Beijing 100049, China}

\author{Yueqiang Gu}
\email{Y. Gu and Y. Xu contributed equally to this work.}
\affiliation{Beijing National Laboratory for Condensed Matter Physics, and Institute of Physics, Chinese Academy of Sciences, Beijing 100190, China}
\affiliation{University of Chinese Academy of Sciences, Beijing 100049, China}

\author{Tiantian Zhang}
\affiliation{Beijing National Laboratory for Condensed Matter Physics, and Institute of Physics, Chinese Academy of Sciences, Beijing 100190, China}
\affiliation{University of Chinese Academy of Sciences, Beijing 100049, China}

\author{Chen Fang}
\affiliation{Beijing National Laboratory for Condensed Matter Physics, and Institute of Physics, Chinese Academy of Sciences, Beijing 100190, China}
\affiliation{Songshan Lake Materials Laboratory, Guangdong 523808, China}
\affiliation{CAS Center for Excellence in Topological Quantum Computation, University of Chinese Academy of Sciences, Beijing 100190, China}

\author{Zhong Fang}
\affiliation{Beijing National Laboratory for Condensed Matter Physics, and Institute of Physics, Chinese Academy of Sciences, Beijing 100190, China}
\affiliation{University of Chinese Academy of Sciences, Beijing 100049, China}

\author {Xian-Lei Sheng}
\email{xlsheng@buaa.edu.cn}
\affiliation{Department of Physics, Key Laboratory of Micro-nano Measurement-Manipulation and Physics (Ministry of Education), Beihang University, Beijing 100191, China}

\author{Hongming Weng}
\email{hmweng@iphy.ac.cn}
\affiliation{Beijing National Laboratory for Condensed Matter Physics, and Institute of Physics, Chinese Academy of Sciences, Beijing 100190, China}
\affiliation{University of Chinese Academy of Sciences, Beijing 100049, China}
\affiliation{Songshan Lake Materials Laboratory, Guangdong 523808, China}
\affiliation{CAS Center for Excellence in Topological Quantum Computation, University of Chinese Academy of Sciences, Beijing 100190, China}

\begin{abstract}
Based on first-principles calculations and effective model analysis, we propose that the noncentrosymmetric superconductor YCoC$_2$ in normal state is a topological semimetal. In the absence of spin-orbit coupling (SOC), it can host  two intersecting nodal rings protected by two mirror planes, respectively.
One ring is composed of type-I nodes, where the two crossing bands have opposite slop sign in their dispersions. The other ring consists of both type-I and type-II nodes (the slop signs of the two bands are the same in certain direction). In the presence of SOC, the former nodal ring is gapped totally while  the later one evolves into ten pairs of Weyl nodes, with two of them being type-I and eight being type-II. The type-II Weyl nodes are further classified into two kinds with different velocity matrix when described in Weyl equation near the nodes. Fermi arcs from topological surface states are observed in the surface projected energy dispersions. It is notably that YCoC$_2$ has been reported as a superconductor with critical temperature $T_c$ of 4.2 K. This makes it very attractive since including superconducting into topological semimetal state might result in topological superconductivity and be used to synthesize Majorana zero modes.

\end{abstract}
\maketitle


\section{Introduction}

Topological quantum states and topological materials have attracted great interests from researchers in both fields of condensed matter physics and materials sciences. The studies on topological insulators have achieved great success.~\cite{Hasan2010,Qi2011,Bansil2016, MRS_review} However, in recent years the research focus has been shifted towards topological semimetals/metals~\cite{Chiu2016,Burkov2016,Yan2017,Armitage2017,TSM_review}. In these materials, the electronic band structure has the feature that there are energy nodes formed by band crossing close enough to the Fermi level and these bands dominate the Fermi surface so that they possess  topologically nontrivial properties.
These energy nodes can be classified into several classes according to their dimensionality, degeneracy and even the velocities of the involved bands in crystal momentum space around the nodes. The crossings in the Weyl and Dirac semimetals are isolated zero-dimensional (0D) nodal points ~\cite{Armitage2017,Wan2011,Murakami2007,Young2012}, in which the conduction and valence bands cross linearly at isolated momentum $k$-points. The quasiparticles around the nodal points are analogy of the massless Weyl/Dirac fermions from the standard model, while they can lead to Fermi arc when the Weyl/Dirac nodes are projected onto a certain surface, which is absent for the particles of Weyl/Dirac fermions. 
In nodal line semimetals, the bands cross along 1D path in the momentum space~\cite{Weng2015,Yang2014,Mullen2015,Yu2015,Kim2015,Chen2015,Fang2016}, and the resulted drumhead surface states can be observed. The band crossings can also form 2D nodal surface, where each nodal point on the surface is a crossing point with linear dispersions along the surface normal direction~\cite{LiangQF2016,WuWK2018,Zhong2016Towards-N,ZhangXM2018}. The topological semimetals with nodes of three-, six- and eight-fold degeneracy have also been proposed~\cite{Bradlyn2016, Weng2016b, Weng2016a, zhu2016}. 

Most of the above topological semimetal states have been realized in specific materials~\cite{Wan2011,XuG2011PRL,WengHM_2015TaAs,Huang2015,Lv2015, Xu2015,Ruan2016ei,Ruan2016PRL,Sheng_TlN,Sheng2017JPCL,ShengQAHE, ChenC2017,ZhangXM2017,PhysRevB.98.201112,arXiv180808608L,LiuPRL2017,GuanS2017PRM,Liu2018,FengBJ2018,LiS2018,FuBT2018,PRB97045131,PRB96081106,tang2019efficient,tang2019topological,WanNature2019, MoP_2017}. For example, Weyl semimetal was initially proposed in magnetic pyrochlore iridates~\cite{Wan2011} and HgCr$_2$Se$_4$~\cite{XuG2011PRL} but neither has been confirmed experimentally. The first experimentally confirmed and widely studied Weyl semimetals are nonmagnetic TaAs family,~\cite{Lv2015, Xu2015} which was predicted by Weng {\it et al.}~\cite{WengHM_2015TaAs} and Huang et al.~\cite{Huang2015}. Dirac fermions has been  proposed and observed in Na$_3$Bi~\cite{WangZJ2012,Liu864} and Cd$_3$As$_2$~\cite{WangZJ2013,Neupane2014kca,Borisenko2014,Liu2014hr}. For nodal line semimetals, they are firstly predicted in carbon network materials with three intersecting nodal rings~\cite{Weng2015,Chen2015}. Chain like nodal line metals have been proposed in IrF$_4$ with Weyl type~\cite{Bzdusek2016} and ReO$_2$ with Dirac type~\cite{Wang2017NC}, both of which are protected by nonsymmorphic symmetries and robust against spin-orbit coupling (SOC).  

Here, we propose that a noncentrosymmetric material YCoC$_2$, which has been reported as a superconductor with a critical temperature of $T_c=4.2$ K by magnetization, resistivity, and heat capacity  measurements~\cite{Cigarroa2014},  is a topological semimetal. In the absence of SOC, it can host topological nodal line state with two intersecting nodal rings. In the presence of SOC, similar to TaAs~\cite{WengHM_2015TaAs} and HfC,~\cite{YuR2017PRL} there opens band gap along the nodal lines but leads to ten pairs of Weyl points nearby the original nodal lines. The obtained ten pairs of Weyl nodes are related by time reversal and mirror symmetries. Only three of them are nonequivalent. Both type-I and type-II band dispersions are observed around these Weyl nodes. The Fermi arc surface states connecting projected Weyl points with opposite chirality have been discussed.

\section{Crystal Structure and Methodology}

\begin{figure}[tbp]
\includegraphics[angle=0,width=1\linewidth]{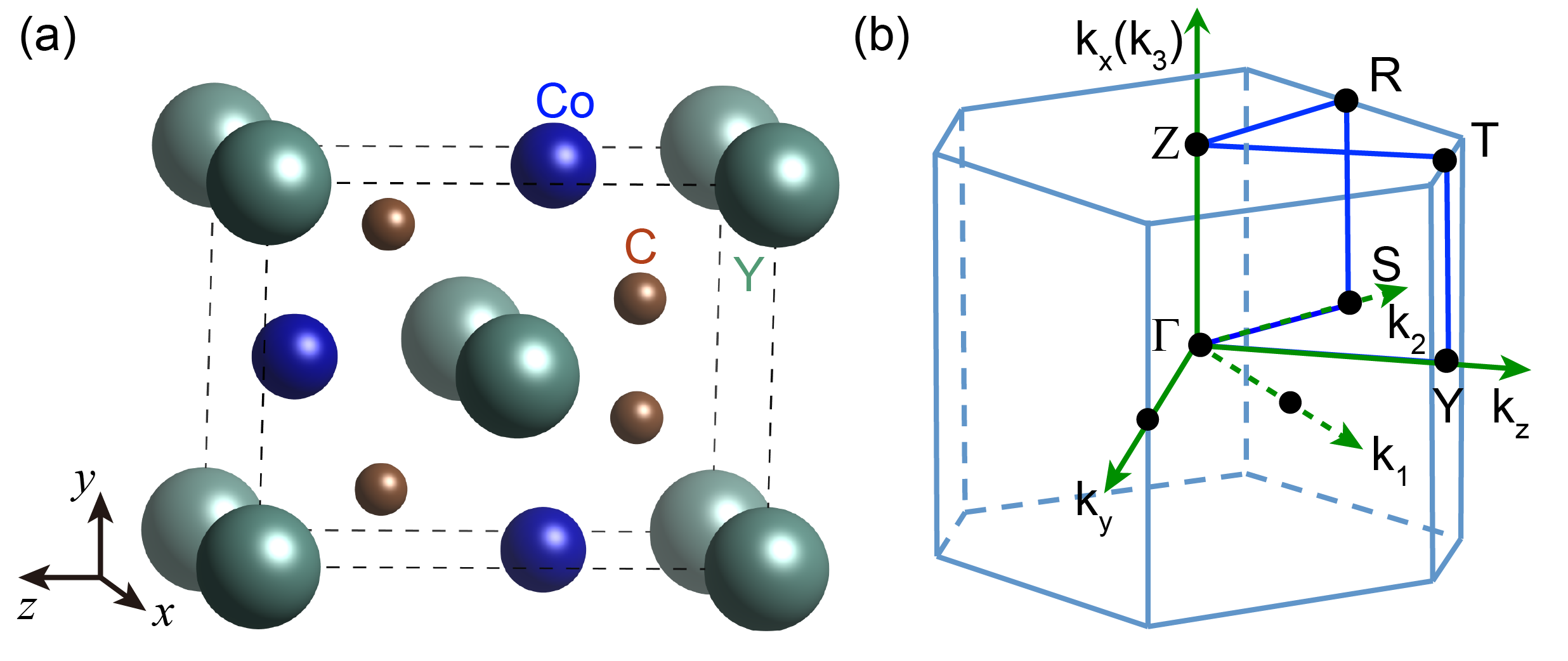}
\caption{(a) Crystalline structure of the base-centered orthorhombic lattice YCoC$_2$ consisting two primitive cells. (b) The corresponding first Brillouin zone of primitive cell together with the Cartesian axes ( $\mathbf{k}_x$, $\mathbf{k}_y$ and $\mathbf{k}_z$) and reciprocal lattice vectors ($\mathbf{k}_1$, $\mathbf{k}_2$ and $\mathbf{k}_3$) with relationship of   $\mathbf{k}_x=\mathbf{k}_3$, $\mathbf{k}_y=\mathbf{k}_1-\mathbf{k}_2$ and $\mathbf{k}_z=\mathbf{k}_1+\mathbf{k}_2$.}
\label{Fig:Crys}
\end{figure}

YCoC$_2$ adopts the base-centered orthorhombic lattice with space group Amm2 (No. 38).  The experimental lattice constants were used in the following calculations with $a$=3.54~\AA, $b$=4.52~\AA~ and $c$=6.03~\AA~\cite{jeitschko1986ternary}. The Y atoms occupy 2a(0, 0, 0) Wyckoff position, Co atoms occupy the 2b(0.5, 0.0, 0.3856) one and C atoms are at 4e(0.5, 0.345, 0.2086) one. The space group consists of two mirror planes $\mathcal{M}_x$ and $\mathcal{M}_y$, as well as a $C_2$ rotation around $z$-axis. Fig.~\ref{Fig:Crys} (a) shows the unit cell of the base-centered orthorhombic lattice, which consists of two primitive cells. The corresponding first Brillouin zone (BZ) with high symmetry points and paths is shown in Fig.~\ref{Fig:Crys}(b). 

The first-principle calculations are based on the density functional theory (DFT), as implemented in the Vienna \emph{ab initio} simulation package~\cite{kresse1993VASP, kresse1996VASP}.  The projector augmented wave method is adopted~\cite{blochl1994PAW}. The generalized gradient approximation (GGA) with the Perdew-Burke-Ernzerhof (PBE) realization~\cite{PBE1996PBE} is taken for the exchange-correlation potential. For all calculations, the experimental lattice structure is used. The BZ sampling is performed by using $k$ grids with a $10 \times 10 \times 10$ mesh within a $\Gamma$-centered sampling scheme. As the transition metal Co 3\emph{d} orbitals may have notable correlation effects, we have checked the $U$ dependence of the results by GGA+$U$ calculations~\cite{Dudarev1998} (see Appendix.~\ref{plusU}). The key features are found to be qualitatively the same as the GGA calculation.

\section{In the Absence of SOC: two intersecting nodal rings}

\begin{figure}[tbp]
\includegraphics[angle=0,width=1\linewidth]{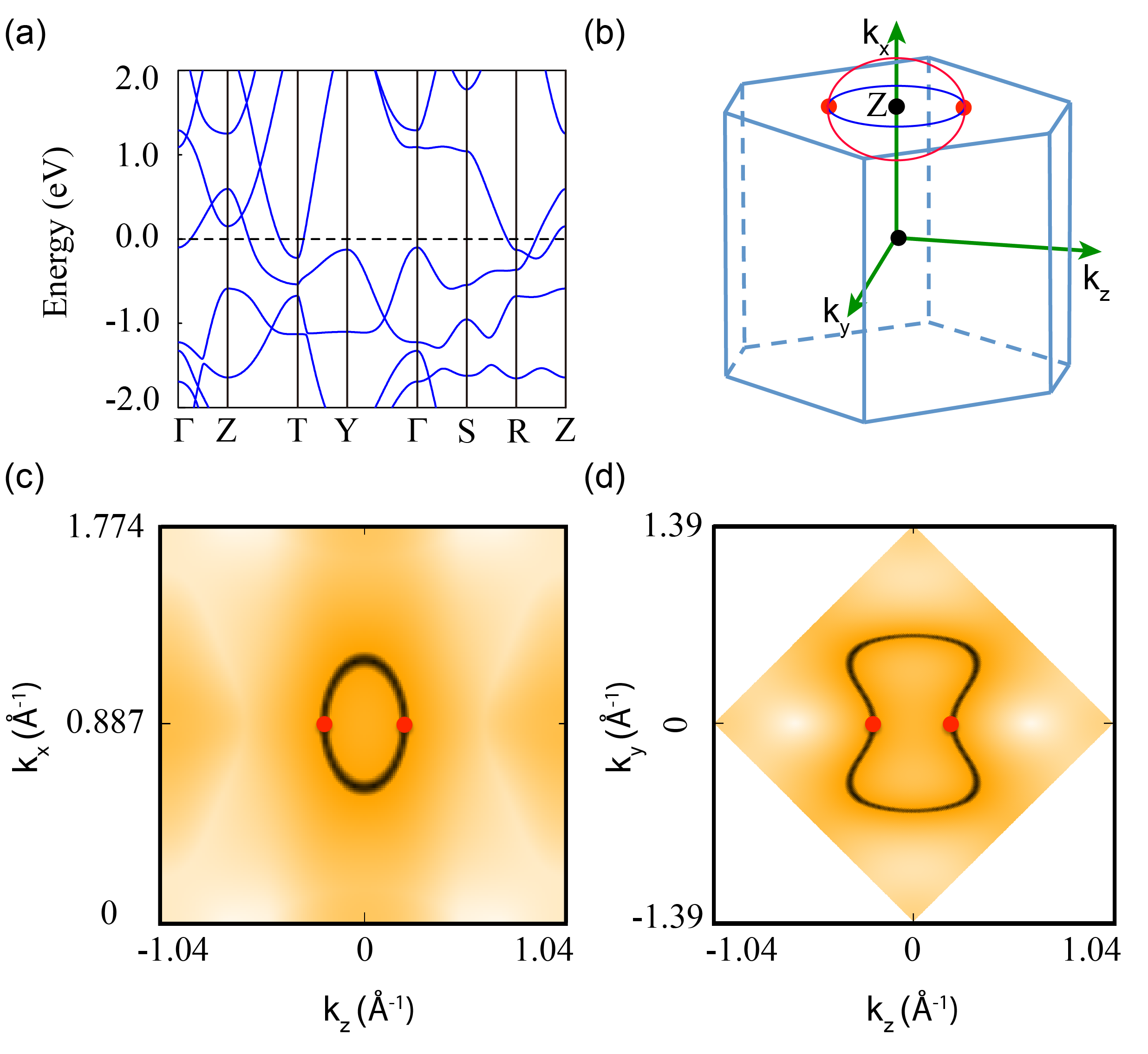}
\caption{(a) Band structure of YCoC$_2$ without SOC. (b) Schematic view of the two nodal rings on the mirror planes. (c) and (d) two nodal rings from first-principles calculations on $k_1=k_2$ and $k_3=\pi$ plane, respectively. The red dots mark the touching points of the two loops.}
\label{Fignosoc}
\end{figure}

In the absence of SOC, YCoC$_2$ is a topological metal with band-crossing dispersions around the high symmetry point Z in BZ around Fermi energy as shown in Fig.~\ref{Fignosoc}(a). The orbital projected bands show that the two crossing bands around Z are dominated by 3$d$-orbitals of Co, especially the $d_{xy}$ and $d_{z^2}$ orbitals for the conduction and valence band, respectively. 
Around T point, there is an electronic pocket from another band. Therefore, there are three bands crossing Fermi level, leading to a large Fermi surface. Intensive scanning calculations in BZ show that two nodal rings exist around the Fermi energy. The two nodal rings lie in the $k_3=\pi$ and the $k_1=k_2$ mirror planes, respectively. The two rings are tangent at two points on the intersection of the mirror plane, forming a X like shape, so named nodal X-ring [see Fig.~\ref{Fignosoc}(b)], as shown in Figs.~\ref{Fignosoc}(c) and (d).

The nodal X-ring is protected by time reversal symmetry and mirror symmetries $\mathcal{M}_x$ and $\mathcal{M}_y$. To see this more clearly, we constructed a $k \cdot p$ low-energy Hamiltonian around Z point. The symmetry here is characterized by $C_{2v}$ point group, which consists of two mirror planes $\mathcal{M}_x: (x, y, z) \to (-x, y, z)$, and $\mathcal{M}_y:(x, y, z) \to (x, -y, z)$. We can construct a minimal low-energy model for the two crossing bands around Z. 
\begin{equation}
\mathcal{H}_Z(\mathbf{k})=\varepsilon_0(\mathbf{k})+d_1(\mathbf{k})\sigma_x + d_2(\mathbf{k})\sigma_y +d_3(\mathbf{k})\sigma_z,
\end{equation}
where $d_i$($\mathbf{k}$) ($i=1,2,3$) are real functions of momentum $\mathbf{k}$ and the vector $\mathbf{k}$ is measured relative to the Z point. The first term is proportional to the identity matrix with a real function $\varepsilon_0(\mathbf{k})$. In the absence of SOC, the time-reversal symmetry operator is represented by $\mathcal{T}=\mathcal{K}$ which is the complex conjugate satisfying $\mathcal{T}^2=1$.  Considering the above constraints, the Hamiltonian should satisfy the following requirements:
\begin{equation}\label{eqT}
\mathcal{T}\mathcal{H}_Z(\mathbf{k})\mathcal{T}^{-1}= \mathcal{H}_Z(-\mathbf{k}),
\end{equation}

\begin{equation}\label{eqMx}
\mathcal{M}_x\mathcal{H}_Z(\mathbf{k})\mathcal{M}_x^{-1}= \mathcal{H}_Z(-k_x, k_y, k_z),
\end{equation}

\begin{equation}\label{eqMy}
\mathcal{M}_y\mathcal{H}_Z(\mathbf{k})\mathcal{M}_y^{-1}= \mathcal{H}_Z(k_x, -k_y, k_z),
\end{equation}
Eq.~(\ref{eqT}) requires that $d_2(\mathbf{k})$ is an odd function of $\mathbf{k}$, while $d_{1,3}(\mathbf{k})$ are  even functions of $\mathbf{k}$. The eigenfunctions of the two crossing bands are also eigenfunctions of mirror symmetries $\mathcal{M}_x$ and $\mathcal{M}_y$. The first-principles calculations show that the irreducible representations of the two crossing bands are different. Thus, the matrix representation of the two mirror operators could be $\sigma_z$. Thus, up to second order, the Hamiltonian reads,
\begin{equation}\label{eqH}
H_Z(\mathbf{k})=\varepsilon_0(\mathbf{k})+\\
\left(
\begin{array}{cc}
d_3(\mathbf{k}) & bk_xk_y \\
bk_xk_y  &  -d_3(\mathbf{k})
\end{array}
\right)\\
\end{equation}
where $\varepsilon_0(\mathbf{k})=a_0+a_1k_x^2+a_2k_y^2+a_3k_z^2$, $d_3(\mathbf{k})=c_0+c_1k_x^2+c_2k_y^2+c_3k_z^2$. The parameters $a_i$ $c_i$ and $b$ can be
derived by fitting the dispersions to those of first-principles calculations. 
The two bands around Z point near the Fermi energy with the inverted structure lead to $c_0 > 0$ and $c_{1,2,3} < 0$, which is important to the existence of nodal rings. On the plane $k_x=0$, eq.~(\ref{eqH}) leads to 
\begin{equation}
c_0+c_2k_y^2+c_3k_z^2=0
\end{equation}
which gives the band-crossing points to form a circle in the $k_y$-$k_z$ plane. Similarly, on the plane $k_y=0$, from eq.~(\ref{eqH}) one gets
\begin{equation}
c_0+c_1k_x^2+c_3k_z^2=0
\end{equation}
which leads to another nodal ring in the $k_x$-$k_z$ plane. 

Based on the $k\cdot p$ model, it can be proved that there is no other band crossing points in the BZ except the nodal X-ring on the two mirror planes. In general, the eigenvalues of eq.~(\ref{eqH}) take the form
\begin{equation}
  E=\varepsilon_0(\mathbf{k}) \pm \sqrt{(d_3(\mathbf{k}))^2 + (bk_xk_y)^2}
\end{equation}
To get band crossing points, both the terms $(d_3(\mathbf{k}))^2$ and $(bk_xk_y)^2$ should be zero. Then, the second term requires that either $k_x=0$ or $k_y=0$. Thus, the nodal points only exist on the two planes. Since the k-points here are measured from $Z$, the two k-planes correspond to $k_3=\pi$ and $k_1=k_2$ planes, respectively. The conclusions have been confirmed by  the first-principles calculations [See Fig.~\ref{Fignosoc}].

The $\epsilon_0(\mathbf{k})$ term not only changes the shape of the nodal rings but also causes the tilt of the cone. It will not affect the existence of nodal rings. In the $k_3=\pi$ plane, a hybrid nodal line emerges when both type-I and type-II dispersions coexist. In the $k_1=k_2$ plane, the nodal ring is of type-I. 

After fitting the parameters, we got $a_0=0.3651$, $a_1=0.6704$, $a_2=0.5019$, $a_3=-2.0301$, $b=44.1078$, $c_0=0.2203$, $c_1=-2.6212$, $c_2=-0.5063$ and $c_3=-5.0572$. The nodal X-ring states are well reproduced and agree well with the first-principles calculations.

\section{In the presence of SOC: Hybrid Weyl semimetal}\label{S3}

\begin{figure}[tbp]
\includegraphics[angle=0,width=1\linewidth]{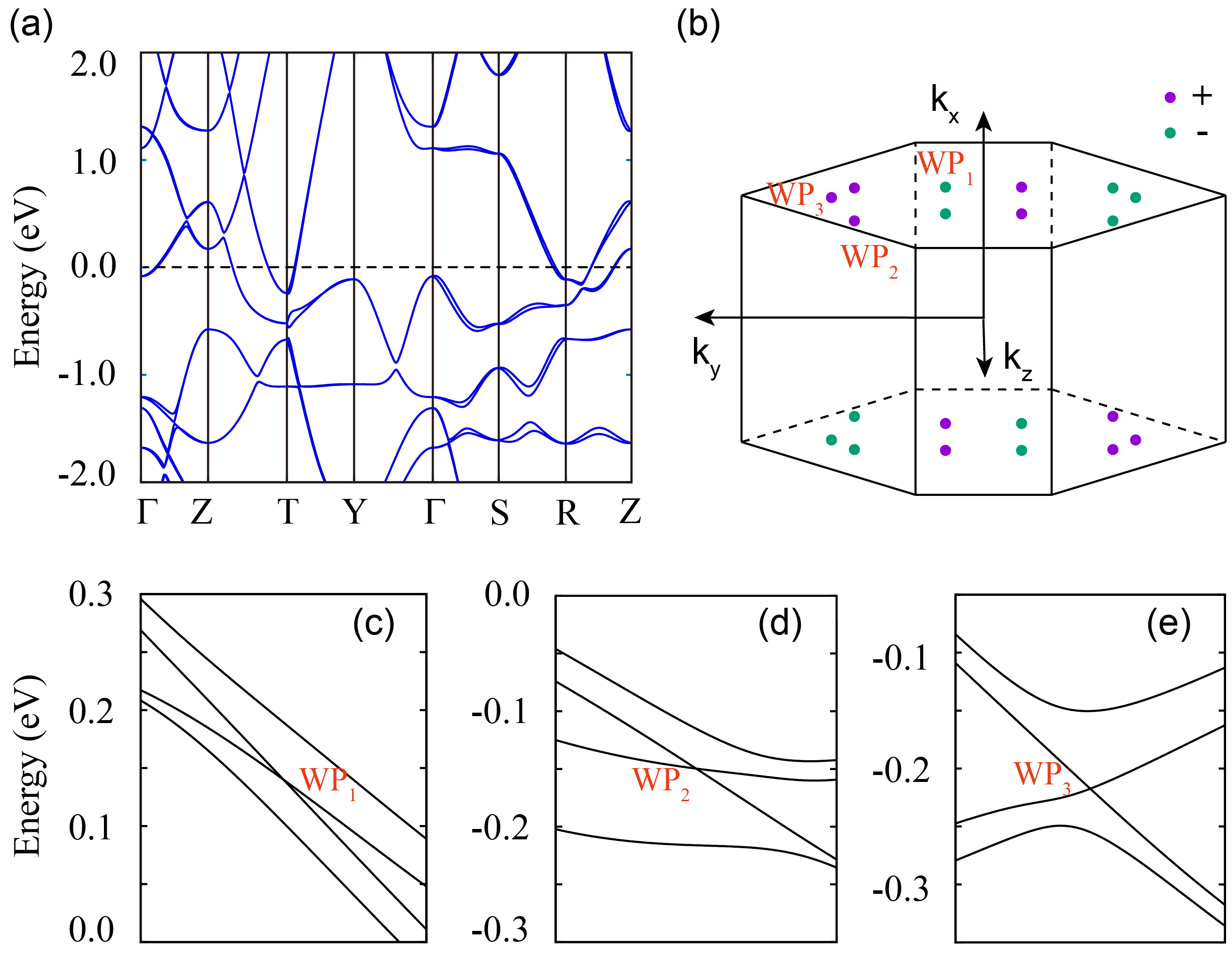}
\caption{(a) Band structure of YCoC$_2$ with SOC. (b) The position and chirality of the ten pairs of Weyl points in first BZ which can be obtained by three nonequivalent Weyl points and the mirror and time reversal symmetry operations. (c) and (d) are the type-II Weyl point with tilted cone dispersion for WP$_1$ and WP$_2$, respectively. (e) Type-I Weyl point with upright cone dispersion for WP$_3$.}
\label{Figsoc}
\end{figure}

\begin{table}[b]
\caption{The Cartesian coordinates and energy of the three nonequivalent Weyl points. }\label{table1}
\begin{centering}
\begin{tabular}{ccc}
\hline \hline
 Weyl point & Position ($\rm \AA^{-1}$ )& Energy (eV)\tabularnewline
\hline
$WP_{1}$ & (0.8519,\; 0.1727,\; 0.2655) & 0.1368 \tabularnewline
$WP_{2}$ & (0.8250,\; 0.5804,\; 0.3265) &-0.1513  \tabularnewline
$WP_{3}$ & (0.8715,\; 0.6826,\; 0.0000) &-0.2174  \tabularnewline
\hline \hline
\end{tabular}
\end{centering}
\end{table}

As discussed above, the nodal lines here is protected by mirror symmetry which is vulnerable against SOC. After considering SOC, the nodal lines may decay into Weyl points or be gapped totally into TI, depending on the strength of SOC.~\cite{WengHM_2015TaAs,YuR2017PRL} From band structure in Fig.~\ref{Figsoc}(a), we find that the band crossing points at high symmetry lines are all gapped. The topological invariants of $Z_2$ indices and the mirror Chern numbers of the two mirror plans have been checked to be zero.~\cite{ZhangNature2019,BernevigNature2019,WanNature2019} These are consistent with the results in our database materiae~\cite{ZhangNature2019} for the topological classification of known nonmagnetic materials. To search the possible band crossing points away from the nodal lines, we have generated atomic like Wannier functions for Y-$4d$, Co-$3d$ and C-$2p$ orbitals by using the Wannier90 package~\cite{souza2001}. The band structures from the tight-binding model based on Wannier functions agree well with the first-principles calculations. We find that there are three kinds of nonequivalent Weyl nodes in ten pairs. All the positions and energies of the three types of Weyl nodes as listed in Table~\ref{table1}.  The chirality of each Weyl node can be determined by integrating the Berry curvature on a sphere enclosing it. The purple and green colors indicate the Weyl point with chirality of $1$ and $-1$, respectively. It should be noted that all the ten pairs of Weyl nodes are close to the nodal ring in $k_3=\pi$ plane. The other nodal ring in $k_1=k_2$ plane is gapped totally, consistent with the analysis in Ref.~\onlinecite{YuR2017PRL}. 

The relativistic Weyl fermions feature an upright Weyl-cone dispersion protected by the particle-hole symmetry. However, particle-hole symmetry is not fundamental in condensed matter and its absence allows the conical band crossings to be tilted. The nodal points can be classified into two types depending on the degree of tilt.  For type-I Weyl points, the cone is slightly tilted and the electron and hole states are separated in energy, while for type-II Weyl nodes, the cone is completely tipped over so that the electron and hole like states coexist at the same energy. Figs.~\ref{Figsoc}(c-e) show the band structures of the three nonequivalent Weyl nodes, marked as $WP_1$, $WP_2$ and $WP_3$. 
For $WP_1$ and $WP_2$, the band structures along some special directions have Weyl cones tilted over as shown in Figs.~\ref{Figsoc} (c) and (d). Thus, $WP_1$ and $WP_2$ are type-II Weyl points. The $WP_3$ is of type-I and its band structure only slightly tilted plotted in Fig.~\ref{Figsoc}(e). 

\begin{figure}[tbp]
\includegraphics[angle=0,width=1\linewidth]{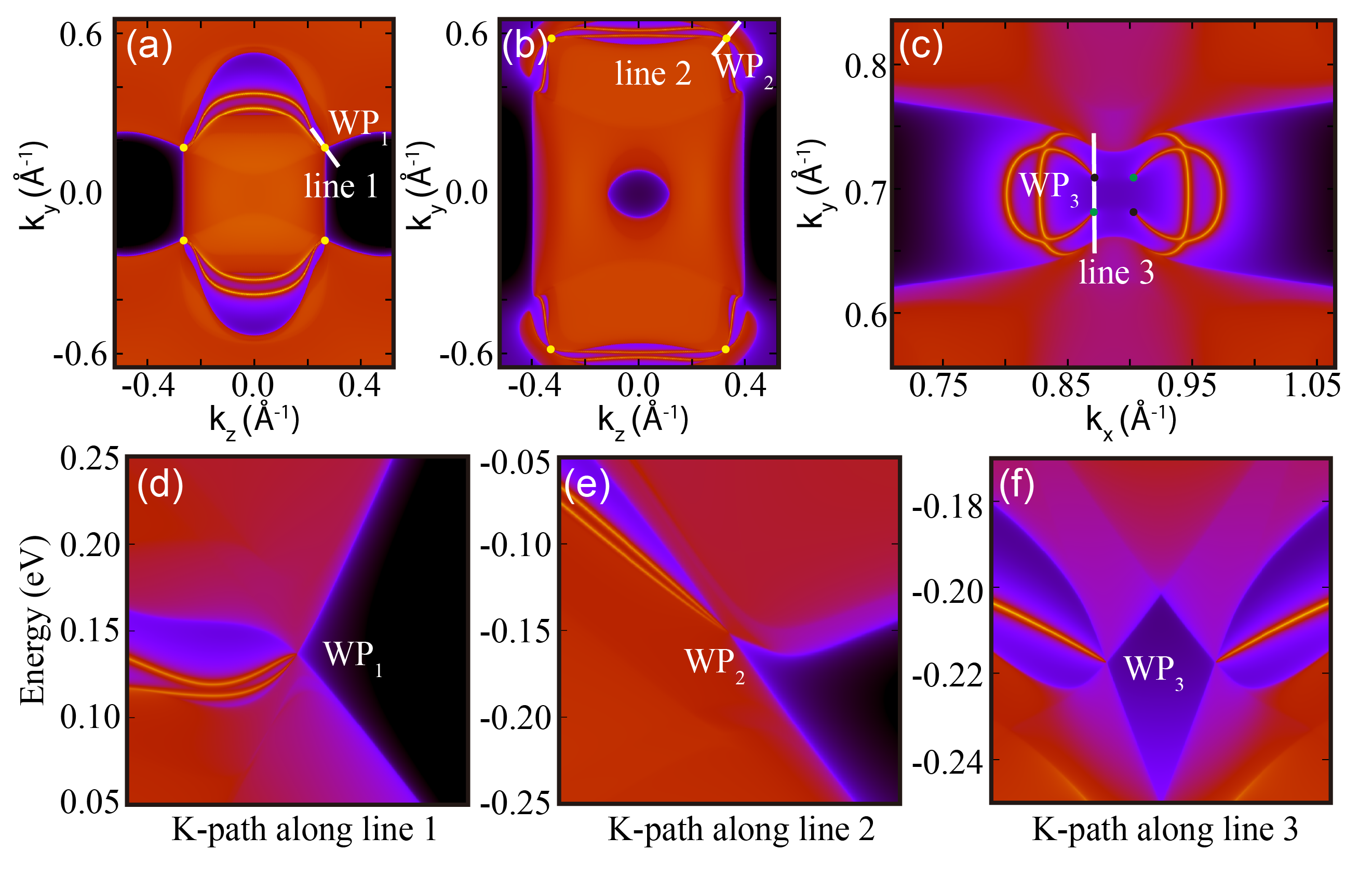}
\caption{Fermi surface for the three kinds of Weyl points at different Fermi level $E_F$. (a) $E_F=0.1368$ eV, (b) $E_F=-0.1513$ eV and (c) $E_F=-0.2174$ eV. (d-f) Topological surface bands along the corresponding path marked in (a-c).}
\label{Fig:surf}
\end{figure}
 
To see the positions of these Weyl nodes clearly, it is necessary to explain the relation between Cartesian coordinates in unit of \AA$^{-1}$ and fractional coordinates in unit of three reciprocal lattice vectors. The Cartesian axes $\mathbf{k}_x$, $\mathbf{k}_y$ and $\mathbf{k}_z$ can be related with reciprocal lattice vectors by  $\mathbf{k}_x=\mathbf{k}_3$, $\mathbf{k}_y=\mathbf{k}_1-\mathbf{k}_2$ and $\mathbf{k}_z=\mathbf{k}_1+\mathbf{k}_2$ (See Fig.~\ref{Fig:Crys} (b)). The Cartesian coordinates and energy of the three nonequivalent Weyl points are list in Table~\ref{table1}. One can see that the WP$_1$ is at $(0.8519,0.1727,0.2655)$, WP$_2$ is at $(0.8250,0.5804,0.3265)$ and WP$_3$ is at $(0.8715,0.6826,0.0000)$. Considering two mirror planes and time reversal symmetries, one gets that there are four pairs of WP$_1$, four pairs of WP$_2$, and two pairs of WP$_3$, respectively. These Weyl points are marked as dots in Fig.~\ref{Figsoc}(b).
 
The surface states of a Weyl semimetal features Fermi arcs connecting the projected Weyl points of opposite chirality. We have calculated the surface state dispersions near the Weyl points with the Green's function method~\cite{WU2017}. On the (001) surface and at the Fermi energy $E_F=0.1368$ eV, the four pairs of of Weyl points are projected at four dots. Each dot consists of two overlapped Weyl nodes and there are two Fermi arcs connecting it with others [See Fig.~\ref{Fig:surf}(a)]. Around one projected point, the topological surface dispersions along the path marked as white line in Fig.~\ref{Fig:surf}(a) are shown in Fig.~\ref{Fig:surf}(d). It is clear that there is a solid cone as the surface projection of Weyl node and two surface states connecting it.  Similar electronic features are observed for WP$_2$ in Figs.~\ref{Fig:surf} (b) and (e). For WP$_3$, on the (110) surface ( $k_x$-$k_y$ plane), the two pairs of Weyl points are projected at four different points. Thus, it is obvious that Fermi arcs connect projections of opposite Weyl nodes in Figs.~\ref{Fig:surf} (c) and (f).

\section{Conclusion}
In summary, based on first-principles calculations and effective model analysis, we propose that a noncentrosymmetric material YCoC$_2$ is an interesting topological material. In the absence of spin-orbit coupling (SOC), it can host a three-dimensional nodal X-ring metal state, consisting of two intersecting nodal loops in two mirror planes ($k_1=k_2$ and $k_3=\pi$). After carefully scanning the dispersions in the two planes, we found that the nodal loop in $k_1=k_2$ is type-I, while in $k_3=\pi$ plane, the band dispersion consist of both type-I and type-II band crossings which leads to a hybrid nodal loop.   In the presence of SOC, the nodal loop in $k_1=k_2$ plane was gapped totally while  the nodal loop in $k_3=\pi$ plane  evolves into Weyl points. There are ten pairs of Weyl points in the first Brillouin zone labeled as three nonequivalent points and connected by mirror and time  , with four pairs of type-I and six pairs of type-II Weyl nodes.  Fermi arc surface states are observed both on (001) and (100) surfaces. This material offers a convenient platform to explore the intriguing physics of nodal X-ring and coexistence of type-I and type-II Weyl fermions. The reported superconductivity with $T_c$=4.2 K in YCoC$_2$ makes it attractive in realizing topological superconductivity and Majorana modes.

\begin{acknowledgements}
The authors thank Q.S. Wu and R. Yu for valuable discussions. XLS was supported by the NSF of China (No. 11504013). Y.X, Y.G. T.Z. C.F. Z.F. and H.W. are supported by National Key Research and Development Program of China (No. 2016YFA0300600 and 2018YFA0305700), the National Natural Science Foundation of China (Grant No. 11674369), the ``Strategic Priority Research Program (B)" of the Chinese Academy of Sciences (Grant Nos. XDB28000000 and XXH13506-202), the Science Challenge Project (TZ2016004), the K. C. Wong Education Foundation (GJTD-2018-01), the Beijing Natural Science Foundation (Z180008), and the Beijing Municipal Science and Technology Commission (Z181100004218001).
\end{acknowledgements}

\ \
\par
\ \
\begin{appendix}
\renewcommand{\theequation}{A\arabic{equation}}
\setcounter{equation}{0}
\renewcommand{\thefigure}{A\arabic{figure}}
\setcounter{figure}{0}
\renewcommand{\thetable}{A\arabic{table}}
\setcounter{table}{0}

\section{Band structure with Hubbard U correction}\label{plusU}

\begin{figure}[tbp]
\includegraphics[angle=0,width=1\linewidth]{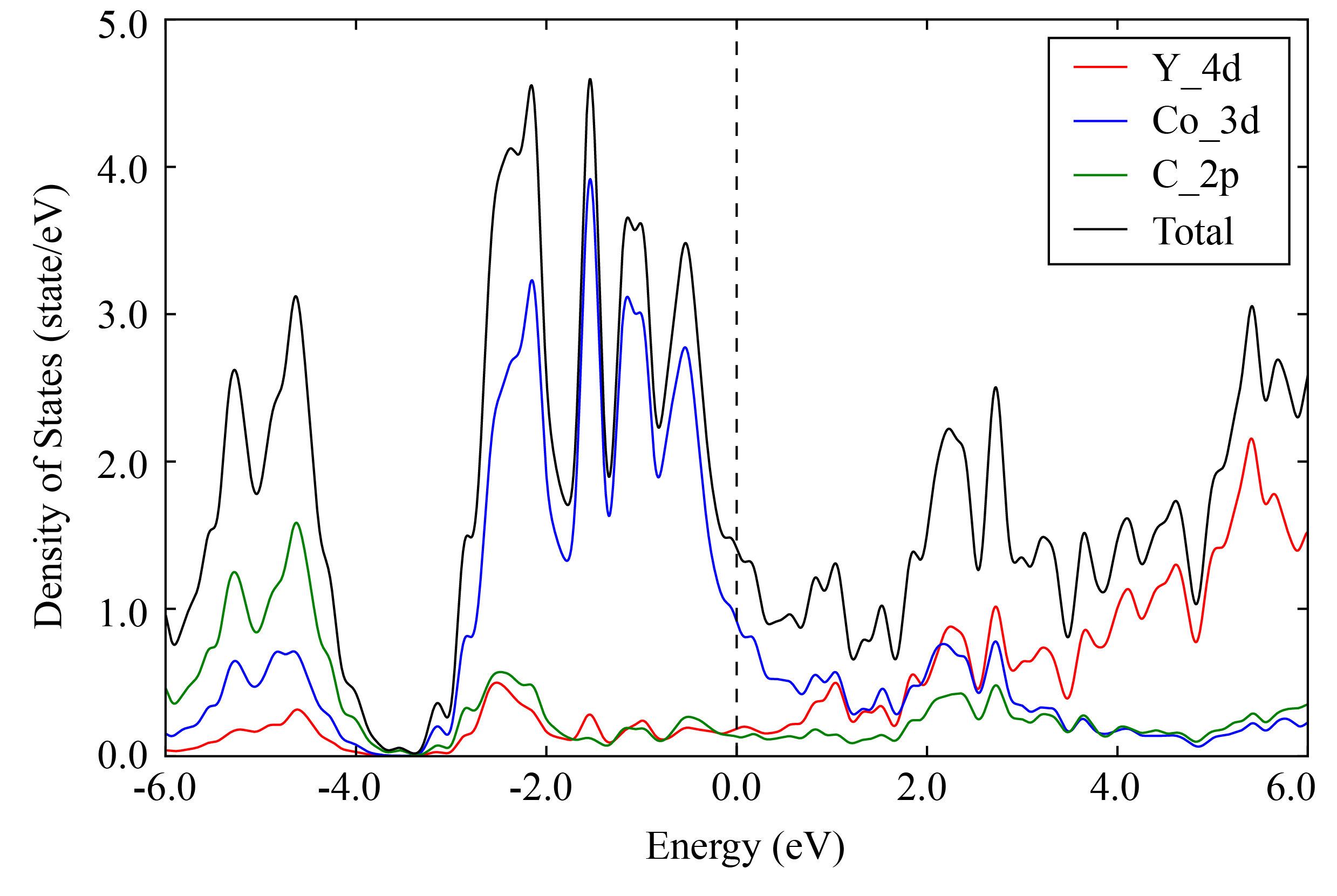}
\caption{Total and partial density of states of YCoC$_2$ indicate that Co-3$d$ states dominate the low energy physics.}
\label{FigDOS}
\end{figure}

\begin{figure}[tbp]
\includegraphics[angle=0,width=1\linewidth]{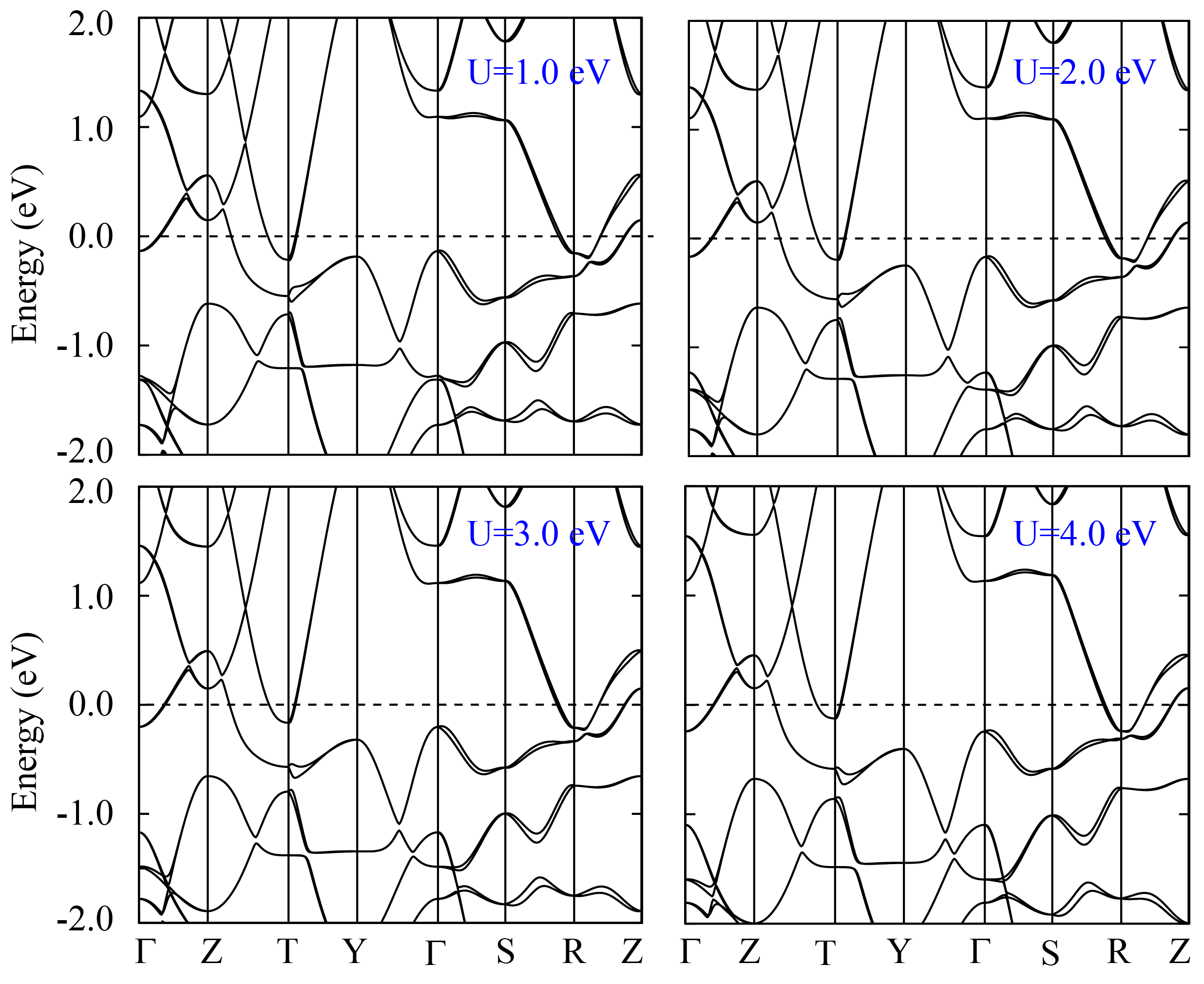}
\caption{Effects of Hubbard $U$ correction. The GGA+$U$+SOC band structure of YCoC$_2$ for $U=1$ eV,  $U=2$ eV,  $U=3$ eV and  $U=4$ eV, showing the same qualitative features with GGA+SOC result.}
\label{FigldaU}
\end{figure}

Since YCoC$_2$ is a transition metal compound, and the low energy physics is dominated by Co-3$d$ orbitals (See Fig.~\ref{FigDOS}), there could be strong electron-electron correlation effects from Co-3$d$ orbitals. Here, we use GGA+$U$+SOC method to calculate the band structure of YCoC$_2$ from $U=0$ to $U=4$ eV. We find that low-energy bands around Fermi level are not sensitive to the $U$ correction (see Fig.~\ref{FigldaU} for results with $U=1$ eV, $U=2$ eV, $U=3$ eV and $U=4$ eV). The first-principles results show qualitatively the same features, and band inversion is maintained, concluding that it is a robust topological Weyl semimetal.

\end{appendix}

\bibliography{NXring_ref}

\end{document}